# Impact of the lead factor of neutron irradiation on the magnetic properties of RPV steels

S. Passanante [a,b,c,*], D. Goijman [b,d], M. R. Neyra Astudillo [c,e], C. D. Anello [f], R. Kempf [g], J. Milano [b,d,h], M. Gómez [c,e], J. Sacanell [a,b]

[a] *Departamento de Física de la Materia Condensada, Gerencia de Investigación y Aplicaciones, Centro Atómico Constituyentes, Comisión Nacional de Energía Atómica, B1650KNA San Martín, Buenos Aires, Argentina.*

[b] *Instituto de Nanociencia y Nanotecnología, Centro Atómico Constituyentes, CNEA-CONICET, B1650KNA San Martín, Buenos Aires, Argentina.*

[c] *Departamento ICES, Gerencia de Desarrollo Tecnológico, Centro Atómico Constituyentes, Comisión Nacional de Energía Atómica, B1650KNA San Martín, Buenos Aires, Argentina.*

[d] *Laboratorio de Resonancias Magnéticas, Comisión Nacional de Energía Atómica, Centro Atómico Bariloche, R8402AGP San Carlos de Bariloche, Argentina.*

[e] *Facultad Regional Delta, Universidad Tecnológica Nacional, San Martín 1171, B2804 Campana, Buenos Aires, Argentina.*

[f] *Gerencia de Materiales, Centro Atómico Constituyentes, Comisión Nacional de Energía Atómica, B1650KNA San Martín, Buenos Aires, Argentina.*

[g] *División Caracterización, GCCN, Centro Atómico Constituyentes, Comisión Nacional de Energía Atómica, B1650KNA San Martín, Buenos Aires, Argentina.*

[h] *Universidad Nacional de Cuyo, Instituto Balseiro, Centro Atómico Bariloche, R8402AGP San Carlos de Bariloche, Argentina.*

Corresponding author. E-mail Address: sebastianpasanante@cnea.gob.ar (S.Passanante)

**Highlights:**

- Magnetic properties of irradiated SA-508 type class 3 steel can be modified with lead factor.
- The coercive field increases with lead factor, while saturation magnetization decreases with lead factor, when the sample is saturated.
- Although AC susceptibility is primarily real, its imaginary component shows changes with the lead factor.
- RMS value of Barkhausen noise increase with lead factor.

Materials used in nuclear reactors are constantly exposed to the effects of neutron irradiation, which leads to changes in their mechanical properties. In particular, the steels employed in reactor pressure vessels experience a reduction in the ductile-to-brittle transition temperature. Given that the pressure vessel is a non-redundant component, understanding this phenomenon is of significant interest.

In this work, we focus on studying the effects of accelerated irradiation by maintaining constant neutron fluence while varying neutron flux, which results in different lead factors. This approach enables the extrapolation of accelerated test results to real operational conditions of the reactor pressure vessels.

In our study, we analyzed irradiated steels using three magnetic techniques. Each technique responds differently to the microstructural changes induced by irradiation, allowing for a better characterization of its effects on the material. Through DC magnetometry, the analysis of minor loops shows that the remanence, coercivity, and saturation of the steel depend not only on the fluence with which the material was irradiated but also on the irradiation rate, which means, more specifically, on the lead factor. The AC susceptibility technique shows that the response of the irradiated steel to the applied magnetic field increases with fluence but also with the lead factor. For the real part of $\chi_{AC}$, there is an increase with the lead factor, while the imaginary component of $\chi_{AC}$ grows with the size of the nanoprecipitates. Finally, Barkhausen noise measurements show a clear increase in the RMS of the signal with the lead factor.

## 1. Introduction

Nuclear energy currently accounts for a significant portion of the global energy supply, recognized for its efficiency, cleanliness, and safety. As energy demands continue to rise, its role is expected to grow even further [1]. However, the generation of nuclear energy requires meticulous control to ensure safety and reliability. This includes rigorous monitoring of the structural components of nuclear reactors, particularly the reactor pressure vessel (RPV), which is a critical element in containing the nuclear reaction and housing essential components such as fuel, coolant, and control rods.

The RPV is one of the most crucial and irreplaceable components of a nuclear reactor, both technically and economically. It is a massive steel structure, up to 15 meters tall and 4 to 6 meters in diameter, with walls between 15 and 30 cm thick. These walls are designed to block neutron radiation effectively, ensuring the safety of operators and the environment [2] [3]. However, prolonged exposure to neutron irradiation leads to gradual degradation of the RPV's material properties by inducing changes in its microstructure [4]. As a result, RPVs are typically designed with an operational lifespan of approximately 40 years. Understanding their structural condition is crucial for planning early decommissioning or safely extending their service life [5].

Two main mechanisms drive RPV degradation: stable matrix damage (SMD) [6] [7], and the formation of irradiation-induced precipitates. Copper-rich precipitates (CRP) are the most common [8], although other elements may also precipitate depending on the steel's composition [9] [10]. SMD occurs when high-energy neutron irradiation disrupts the steel's crystal lattice, forming defects such as vacancies, interstitials, and clusters that degrade its mechanical performance [5] [11]. Simultaneously, precipitates form as certain elements segregate within the ferritic matrix under irradiation, further increasing embrittlement [12]. While CRPs are extensively studied, precipitates involving phosphorus, nickel, silicon, or manganese also contribute to microstructural degradation [13].

Traditionally, the condition of RPVs is assessed through surveillance programs involving the extraction of irradiated Charpy specimens produced from the same steel as the RPV. These specimens are analyzed during reactor shutdowns (scheduled or unscheduled), as accessing them requires halting operations. This approach has notable limitations: it is not continuous, involves handling radioactive material, and relies on destructive testing, meaning samples cannot be reused. Despite these challenges, the Charpy test remains the primary method for determining the ductile-to-brittle transition temperature (DBTT), a critical parameter for RPV safety. The DBTT increases with irradiation, and operating the reactor below this temperature places the RPV in a brittle state, which could lead to catastrophic failure.

In this context, non-destructive testing (NDT) methods can be a promising alternative [14]. Among the most commonly used techniques, ultrasound based techniques [15], as well as those related with optical inspections [16], electrical [17], and magnetic measurements [18] are the most relevant. NDT methods present several advantages, as for example their low cost, and repeatability. An additional advantage of electrical and magnetic measurements is that, once understood, an in-situ device can be designed to monitor them in operation. However, for these methods to be effective, the correlation between measured properties the materials mechanical properties, must be properly understood.

Under experimental conditions, test specimens are irradiated at accelerated rates, achieving fluences (number of neutrons per unit area) equivalent to decades of reactor activity within days or months. This approach allows the study of specific steel properties as a function of fluence within a timeframe suitable for scientific research, providing insights on how the material evolves under irradiation. However, the damage experienced by these samples is not directly equivalent to that of the reactor wall, due to the accelerated nature of the process.

The lead factor (LF) quantifies the degree of acceleration in the embrittlement process. Mathematically, it is defined as the ratio between the fluence on the Charpy specimen and the fluence on the inner wall of the RPV. This parameter is crucial because irradiation conditions, particularly neutron flux, can significantly influence the steel's microstructure, leading to deviations from the behavior observed under normal reactor operating conditions. For instance, under accelerated irradiation (higher flux), the microstructure of the steel may evolve differently, altering its evaluated properties [19]. In practical terms, these differences can introduce discrepancies in predictive models of material property changes, potentially underestimating or overestimating irradiation effects [11]. These observations underscore the importance of considering the LF when interpreting experimental results and extrapolating them to real reactor conditions [20].

In this work, we investigate how the magnetic properties of SA508 type 3 steel are affected by irradiation, with particular emphasis on the role of the lead factor. The magnetic techniques evaluated are non-destructive, highly repeatable, cost-effective, and capable of continuous measurements, making them promising candidates for in-situ monitoring during reactor operation. This is feasible because SA508 steel is ferromagnetic at room temperature, and the CRP formed during irradiation, due to their nanometric size, interact with the movement of magnetic domain walls [19].

We employed three distinct techniques to investigate this phenomenon: We measured DC magnetization with a vibrating sample magnetometer (VSM), AC susceptibility, and Barkhausen noise. The results demonstrate that each technique exhibits sensitivity to different aspects of the structural changes in the steel. Moreover, the sensitivity of these techniques is influenced by the LF, highlighting its role in the interpretation of the effects of irradiation on the material.

By using a VSM, we performed "minor loops" [18] [21], in order to assess the effects of irradiation on the magnetic properties. The procedure consists on measuring several hysteresis loops (M vs H) in which the maximum applied magnetic field for each cycle ($H_{max}$) is increased from a relative low value (much lower than that needed to saturate M) from one loop to another. Each of these curves is called a minor loop, and the characteristic values of the curve are extracted from them. When $H_{max}$ finally reaches the saturation, the resulting cycle is called a major loop.

The analysis of minor loops reveals the influence of irradiation on the microstructure of the material through its magnetic properties [22]. As mentioned before, neutron irradiation provokes the appearance of different defects, such as SMD and CRP. As $H_{max}$ increases, magnetic domain walls are compelled to move and overcome barriers set by these defects [23], which in turn affects the hysteresis parameters, namely the coercive field ($H_c^*$), the remanent magnetization ($M_R^*$), and the maximum magnetization ($M_{max}$). Note that, for a particular minor loop, $M_{max}$ is not the saturation of magnetization.

This makes it a useful (albeit indirect) way to quantify how neutron irradiation affects the steel, bringing it closer to the risk of embrittlement. Additionally, this approach works in both open-loop configurations (using VSM or SQUID systems) that can achieve high magnetic fields, and closed-loop configuration where lower fields are applied.

A complementary strategy, explored in a recent study [24], involves analyzing the magnetization work. This technique focuses only on the major loop, where the applied field (H) reaches the sample's saturation. From this loop, the work required to magnetize the sample during a hysteresis cycle, referred to as $W_{ext}$, is calculated. To minimize the influence of the sample shape on their study, the authors used the same geometry for all, ensuring a similar demagnetizing field effect. The results revealed a clear dependence of $W_{ext}$ on fluence, highlighting its potential as a complementary tool for analyzing irradiation effects on the material.

Another technique used in this work is AC magnetic susceptibility. While commonly applied to study magnetic states such as spin glass [25] or antiferromagnetism (AFM) [26] through magnetization curves as a function of temperature (M(T)), here it is employed at room temperature to analyze how irradiation-induced changes affect magnetic domain walls dynamics. In this technique, the real (Re(χ)) and imaginary (Im(χ)) components of AC susceptibility are measured as a function of the frequency (*f*) of the applied magnetic field to investigate domain wall behavior in irradiated steels. The real component corresponds to the in-phase response of magnetization to the applied field, providing insights into domain wall mobility, while the imaginary component reflects energy dissipation during domain wall motion [27]. A peak in Im(χ) at a specific frequency indicates resonance in domain wall movement, allowing the estimation of the characteristic relaxation time of the process [27]. By analyzing AC susceptibility as a function of *f*, the experiment reveals how the domain walls interact with the precipitates, influencing their mobility and energy dissipation.

Additionally, magnetic Barkhausen noise (BN) was measured. The BN phenomenon occurs in ferromagnetic materials subjected to a varying magnetic field, generating a characteristic crackling or murmuring sound [28]. BN results from abrupt magnetization changes caused by the sudden motion and reorientation of domain walls as they overcome energy barriers such as microstructural defects.

The analysis of BN is widely used in non-destructive testing and materials characterization, providing insights into including grain size, defect density, residual stresses, and hardness [29]. Its key advantages include the fact that is a relatively fast experiment, requires minimal surface preparation and is sensitive to surface effects up to depths of approximately 0.1 mm [30]. In the case of steels used in nuclear reactors, BN is particularly useful for evaluating the effects of irradiation-induced defects, such as those generated by the CRP or SMD [31].

In this study, we present a combination of the previously described these techniques in order to assess the impact of irradiation on the material's microstructure and magnetic properties. The results provide a detailed characterization of irradiation-induced changes, offering valuable insights for future applications in irradiated environments.

## 2. Materials and methods

The study was performed in ASME SA-508 type class 3 steel, which is used in the construction of the nuclear reactor of the Atucha II nuclear plant. In Table 1 we present its detailed composition. The irradiation was carried out in the RA-1 reactor (CNEA, Argentina), which is equipped with a facility for irradiating Charpy specimens. Further information on the irradiation process can be found in reference [32].

Table 2 presents the irradiation conditions. Two of the samples were irradiated, while the third (LF 0) remained unirradiated and was used as a control/reference sample. Since the steel has a high concentration of Cu, the formation of CRP is the predominant mechanism responsible for steel embrittlement [5] [13].

| C | Mn | P | S | Si | Ni | Cr | Mo | V | Cu | Al | Ta | Fe |
|---|---|---|---|---|---|---|---|---|---|---|---|---|
| 0.25 | 1.40 | 0.012 | 0.015 | 0.15 | 0.74 | 0.2 | 0.53 | 0.03 | 0.1 | 0.05 | 0.03 | rest |

Table 1: SA-508 class 3 steel composition. Thermal treatment: Normalizing at 920 °C for 7 h, tempering at 660 °C for 7 h, quenching at 885 °C for 6 h 25 min, followed by tempering at 640 °C for 7 h 50 min. **Note:** Water temperature before quenching: 7 °C; after quenching: 27 °C. [32].

| LF | Fast Flux [n/m² s] | Fluence [n/m²] | Reactor Power [kW] | Irradiation time [hours] |
|---|---|---|---|---|
| 0 | 0 | 0 | 0 | 0 |
| 93 | 1.857 x 10$^{15}$ | 6.6 x 10$^{21}$ | 20 | 984 |
| 186 | 3.715 x 10$^{15}$ | 6.6 x 10$^{21}$ | 40 | 492 |

Table 2: Irradiation conditions of the samples [32].

The magnetization versus magnetic field measurements were performed in a VSM module on the PPMS Versalab™ system by Quantum Design. The system also includes a module to measure magnetic AC susceptibility. The samples were prepared in the form of discs of 3 mm of diameter and thicknesses ranging between 0.07 mm and 0.13 mm, and the magnetization was measured along the sample diameter. The demagnetizing factor was calculated using the approximation of an oblate spheroid [33], and the effective magnetic field was determined as $H_{ef} = H - N_d M$. For AC susceptibility, the correction for real and imaginary component of the AC susceptibility has been applied following the reference [27].

Barkhausen noise (BN) was measured using a custom-built device consisting of a yoke with a 60-turn coil. The maximum current used to induce the magnetic field was 0.5 A. The Barkhausen signal was connected to an amplifier with a low-noise passband ranging from 1 to 500 kHz. For these experiments, we used half of the sample left after the Charpy test.

## 3. Results and Discussion

*3.1. Magnetometry*

In a previous work [32], a significant increase in CRP between the unirradiated sample (LF 0) and the irradiated samples (LF 93 and LF 186) was observed using transmission electron microscopy (TEM). In the LF 93 sample, the CRP are larger (~9 nm) and more widely spaced (~25 nm), whereas in the LF 186 sample, much smaller precipitates (~5 nm) are present, with shorter distances between them (~7 nm). These distributions are crucial for understanding the magnetic properties of the studied samples.

Figure 1(a) shows the coercive field ($H_c^*$) as a function of the maximum applied magnetic field ($H_{max}$) during each minor loop. In the overall magnetic field range, an increase of $H_c^*$ is observed due to the interaction between the magnetic domain walls and the CRP, which act as physical barriers, increasing resistance to domain wall displacement [30] [11] as LF increases, and consequently as the number of CRP grow. Three distinct regions can be identified:

- Low-field region ($H_{max}$ < 200 Oe): $H_c^*$ progressively increases with $H_{max}$. This behavior is interpreted as the initial motion of magnetic domain walls in all samples. As the domains begin to grow, the domain walls encounter their first obstacles, which gradually hinders their movement. The more obstacles the walls face, the higher $H_c^*$ becomes. In this region, the maximum difference in $H_c^*$ between irradiated and non-irradiated samples reaches 8.7% for LF 186.
- Intermediate region (200 Oe < $H_{max}$ < 1000 Oe): $H_c^*$ reaches a plateau, with a maximum near $H_{max}$ ≈ 800 Oe. This maximum reflects a transition where domain walls begin to overcome an increasing number of CRP, though not all barriers have been completely surpassed. In irradiated samples, the maximum of $H_c^*$ shifts to slightly higher $H_{max}$ values, indicating that the increased number of CRP generated by irradiation modifies the interaction with domain walls. Interestingly, the $H_{max}$ corresponding to this maximum remains almost independent of LF, suggesting that this behavior is more closely related to the intrinsic nature of the CRP rather than their density. This maximum could serve as a key feature for monitoring material evolution during reactor operation, though further studies are needed to confirm this hypothesis.

- High-field region ($H_{max}$ > 1000 Oe): $H_c^*$ progressively decreases as domain walls overcome most of the obstacles posed by the CRP and move more freely, reducing resistance to their displacement. This behavior is likely related to the progressive alignment of magnetic domains with the applied field as the material approaches full magnetic saturation.

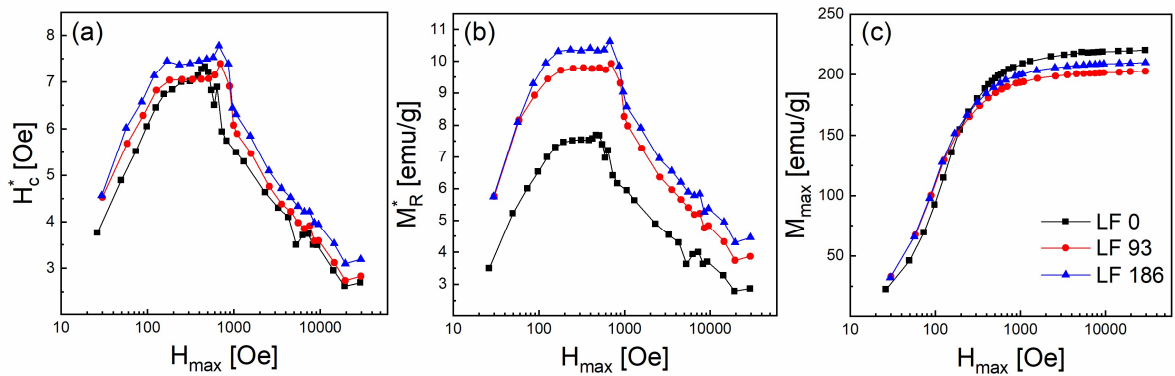

Figure 1:(a) Coercive field ($H_c$), (b) Remanent magnetization ($M_{rem}$), (c) Maximum magnetization ($M_{max}$) as a function of $H_{max}$ for LF 0, LF 93, and LF 186.

Remanent magnetization ($M_{rem}^*$), figure 1(b), follows a similar trend as that displayed by $H_c^*$, with an initial increase, a maximum in the intermediate range and a decrease at high $H_{max}$ values. However, the differences between the curves corresponding to different LF values are more significant, as compared to those observed in $H_c^*$, indicating that $M_{rem}^*$ is even more sensitive to the microstructural changes induced by irradiation [30].

In particular, the difference in $M_R^*$ between irradiated and unirradiated samples can reach 25–35% for LF 93, while for LF 186, this difference increases to the range of 35–55%. The largest differences are observed at high $H_{max}$ fields. This behavior can be explained by the cumulative effect of CRP in irradiated material [21]. As LF increases, CRP act as anchoring points for domain wall movement during demagnetization, making it more difficult for domain walls to fully move, resulting in higher remanent magnetization. This behavior reflects how microstructural defects introduced by irradiation, particularly the density and distribution of CRP, affect the material's ability to retain magnetization after H is removed.

This result reinforces the importance of $M_R^*$ as a key parameter for evaluating the effects of irradiation and the lead factor on the microstructure of steels. The high sensitivity of $M_R^*$ to the density and distribution of CRP positions it as a promising parameter for monitoring irradiation-induced changes under reactor operating conditions.

In the low magnetic field region ($H_{max}$ < 200 Oe), the maximum magnetization ($M_{max}$) shows a similar increase than that showed by $H_c^*$ and $M_R^*$ vs $H_{max}$. From $H_{max}$ ~ 200 Oe to the maximum $H_{max}$, $M_{max}$ stabilizes. Besides, it is clear that $M_{max}$ decreases for the irradiated samples.

The lower $M_{max}$ observed in irradiated samples could be explained through the presence of CRP. As the total volume of CRP grows, the ferritic matrix fraction shrinks, and the contribution of this matrix to the saturation is lower. From these observations, we cannot conclude that CRP act as nonmagnetic zones or have a lower saturation than the ferritic matrix.

In most studies analyzing irradiated steels using minor loops [18] [21], the focus is on the dependence of coefficients from the relationships:

$$W_F^* = W_F^0 \left(\frac{M_{max}}{M_{sat}}\right)^{n_F} \quad W_R^* = W_R^0 \left(\frac{M_R^*}{M_R}\right)^{n_R} \quad H_C^* = H_C^0 \left(\frac{M_R^*}{M_R}\right)^{n_C} \qquad (1)$$

where the values marked with * are derived from minor loops, $M_{sat}$ is the saturation magnetization, and $M_R$ is the remanence value, both obtained from the major loop. It has been demonstrated that initial coefficients, such as $W_R^0$, $W_F^0$, and $H_c^0$, can be correlated with changes in the material's mechanical properties as a function of neutron irradiation fluence [18]. In our case, the fluence was kept constant the flux was varied in the LF 93 and LF 186 samples. We found that the exponents associated with these relations are independent of the LF, showing similar values for the different samples. For the exponents, we obtained $n_F = 1.48$, $n_R = 1.97$, and $n_C = 0.97$. However, a dependence with LF has been observed for the $W_R^0$ and $H_c^0$ coefficients, as shown in Figure 2. For both $W_R^0$ and $H_c^0$ an increase is observed with LF. However, $W_F^0$ seems to be independent of neutron irradiation (~7000 erg/g in all samples).

Among these three coefficients, $W_R^0$ exhibits the most significant change between the irradiated and unirradiated material. This makes $W_R^0$ as a key parameter for characterizing the effects of irradiation, and this is why it is one of the most frequently reported coefficients in the literature.

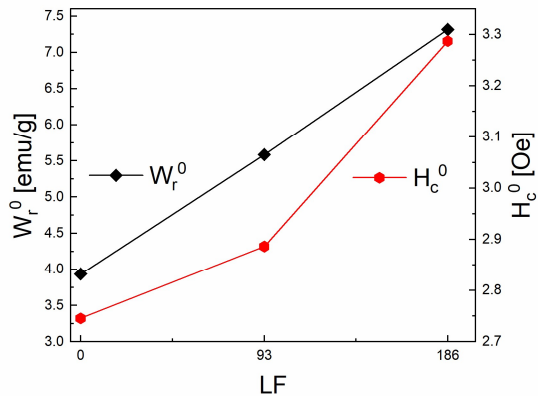

Figure 2: Coefficients $W_R^0$ and $H_c^0$ from minor loops as a function of LF, showing a clear dependence on the lead factor.

To complete the characterization through DC magnetometry, we followed the approach proposed by J. Wang et al. [24]. In this method, the external work of the major hysteresis loop ($W_{ext}$) is analyzed as a function of normalized magnetization $\lambda = \frac{M}{M_{sat}}$, the demagnetizing factor $N_d$, and the irradiation fluence. Wang defines $W_{ext}$ as:

$$W_{ext} = \int_0^M H_{eff} dM + \frac{1}{2} N_d M^2 \qquad (2)$$

Where the first term corresponds to the intrinsic magnetization work, and the second term is the demagnetization energy.

Figure 3 shows the results of $W_{ext}$ versus $\lambda$. For small values of $\lambda$, there is no clear difference between irradiated and unirradiated samples. As $\lambda$ increases, unirradiated sample shows the largest $W_{ext}$. For the irradiated samples, difference between them are observed with LF 186 having a larger $W_{ext}$.

As $W_{ext}$ is calculated from the major loop, and is related to the area enclosed by the M(H) curve, its dependence can be attributed to two different factors: $M_{sat}$, which decreases with LF, and $H_c$, which increases with LF. For this steel, the reduction of $M_{sat}$ is more important than the increase in $H_c$, which leads to a decrease in $W_{ext}$.

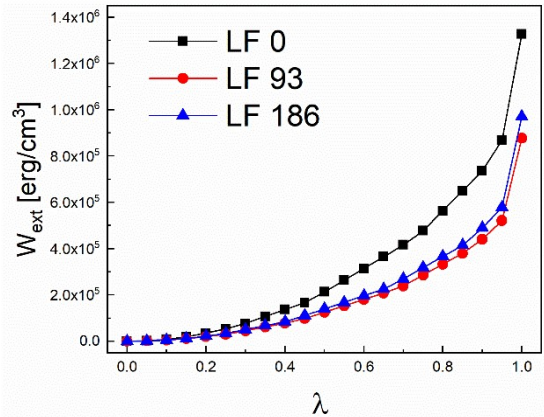

Figure 3: External work of the major loop ($W_{ext}$) as a function of λ (a) and LF (b). Wext decreases with LF, particularly at high λ values.

Finally, as a summary, DC magnetometry shows that irradiation modifies the magnetic response of SA508 and, in particular, that these quantities are sensitive to the lead factor. As $H_{max}$ increases, different regions are observed in $H_c^*$, $M_{rem}^*$, and $M_{max}$. These regions reflect how magnetic domain walls interact with CRP during their motion. At low fields, CRP hinder wall motion and both $H_c^*$ and $M_{rem}^*$ increase. At intermediate fields, a quasi-plateau appears: walls have already overcome the lowest-energy barriers, while higher-energy CRP still pin them. At higher fields (approaching saturation), walls effectively overcome the remaining obstacles and $H_c^*$ and $M_{rem}^*$ decrease with further increases of $H_{max}$. In parallel, the major-loop external work $W_{ext}$ conveys the same picture in terms of the saturation fraction λ: near saturation, $W_{ext}$ decreases with lead factor because the reduction of $M_{max}$ in irradiated samples outweighs the lead-factor-induced increase in coercivity. This behavior is consistent with the CRP size/spacing evolution with LF. We also followed the minor loops relationships approach, which is the most common approach to study irradiated steels in DC magnetometry, and obtained that the exponent $n_c$, $n_f$, and $n_R$ are independent from the LF, same as the $W_F^0$. In contrast, the $W_R^0$ and $H_c^0$ coefficients are sensitive to the change in LF, indicating that irradiation predominantly raises the irreversible work and the effective coercivity associated with domain-wall pinning. Altogether, these VSM-derived metrics provide LF-dependent, microstructure-sensitive markers that are consistent with the observed CRP distribution.

*3.2. AC magnetic susceptibility*

The real (Re(χ)) and imaginary (Im(χ)) components of AC magnetic susceptibility were measured as a function of frequency (f) to analyze the dynamics of domain walls in the irradiated steels. Physically, the experiment involves inducing domain wall motion in response to the frequency of the applied alternating magnetic field. Re(χ) is related to the movement of the walls in-phase with the applied magnetic field, while Im(χ) is associated with energy dissipation during domain wall movement, particularly for irreversible or out-of-phase movements [27].

In Figure 4(a), (b) and (c) we plot Re(χ), Im(χ) and their quotient tan(δ), respectively. We see that Re(χ) is predominant due to the ferromagnetic character of the material. At low frequencies, Re(χ) decreases until it reaches a minimum between 400 and 500 Hz. Beyond this frequency, it abruptly increases and then passes to a near plateau behavior for f > 3000 Hz. Although the overall behavior is similar for all three samples, the irradiated samples exhibit higher Re(χ) values, with this difference becoming more pronounced at higher lead factors. This indicates that irradiation facilitates domain wall movement, possibly due to the formation of smaller domains, as a consequence of the presence of CRP, that are easier to interact with at low frequencies. Another possible explanation, based on our observations in $M_{max}$ for minor loops (Figure 1(c)), is that CRP are more easily magnetized at low fields than the ferritic matrix. As a result, their contribution to the real part of the susceptibility is greater, leading to the observed increase in Re(χ) in irradiated samples.

Although much smaller than the real component, the imaginary component provides information about energy dissipation within the material. As shown in Figure 4(b), Im(χ) shows a similar pattern for all three samples: an initial increase at low frequencies (below 200 Hz), followed by a peak, and then a decrease until around 2000 Hz. At higher frequencies, Im(χ) begins to rise linearly once again.

Irradiated samples display larger Im(χ) values, thus indicating lesser dissipation in the non-irradiated sample. If we look closely to the irradiated samples for f > 2000 Hz, dissipation seems to be slightly larger for LF 93 than for LF 186. This behavior can be related to the size and distribution of obstacles, such as clusters of precipitates in both samples, which affects energy dissipation.

The linear behavior of Im(χ) at high frequencies can be attributed to parasitic currents (eddy currents), a phenomenon also observed in superconductors [34]. The slope obtained for this linear regime is similar for all three cases; however, the y-intercept varies with LF, indicating a dependence on the CRP distribution. After subtracting the linear components from Im(χ) across the entire frequency range, the resulting curves for the three LF values become similar (see SI Fig. 1), suggesting that the remaining signal corresponds to the susceptibility of the ferritic matrix.

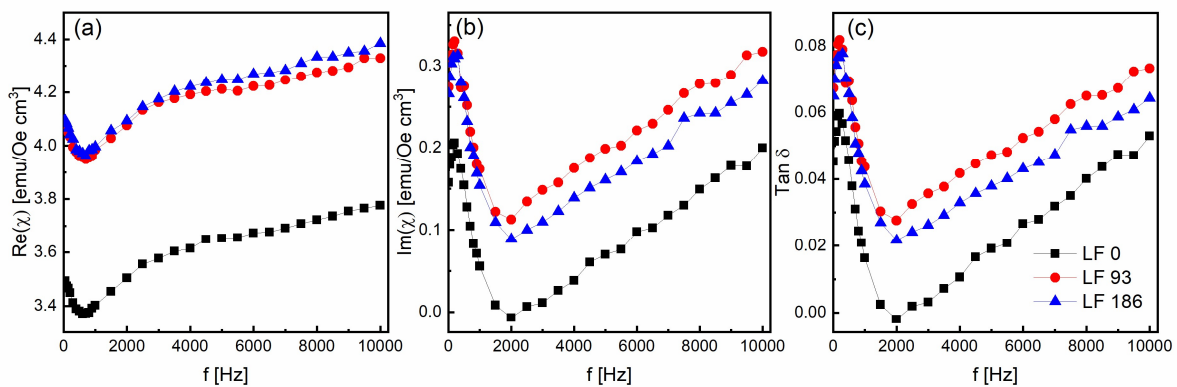

Figure 4: Dependence of AC susceptibility on frequency for LF 0, LF 93, and LF 186: (a) Real component (Re(χ)), (b) Imaginary component (Im(χ)), and (c) tan(δ) ratio.

Figure 4(c) shows tan(δ) (Im(χ)/Re(χ)), where δ is the phase difference between the applied magnetic field and the movement of domain walls. We found a similar behavior in tan(δ) to Im(χ).

For high frequencies, the slope of tan(δ) is proportional to the average size of the magnetic domains [35]. In our case, we found that the non-irradiated sample presents a slope 25 % higher than the two irradiated samples. This indicates that the CRP hinders the formation of larger magnetic domains. As a result, the irradiated samples exhibit a reduction in the size of their magnetic domains.

The AC susceptibility measurements show a clear dependence on irradiation and lead factor. The observed changes can be attributed to the formation of smaller magnetic domains due to the formation of CRP, as compared with the non-irradiated sample. As these magnetic domains are smaller, they become more susceptible to the application of an external magnetic field. Domain-wall motion is principally reversible, even despite the apparition of CRP. However, an increment in the Im(χ) component is found at low frequencies, which is associated with a higher energy dissipation during domain-wall motion.

### 3.3. Barkhausen Noise

The results obtained in Barkhausen noise (BN) measurements show a clear trend between the RMS value of the noise signal and LF (Figure 5). This behavior indicates that LF directly influences the BN response in SA508 steel, as defect density governs changes in magnetic properties [29]. In Barkhausen noise, each peak reflects a discrete domain wall motion event released from pinning sites. Therefore, an increase in defect density could, in principle, be expected to produce a higher number of peaks, as additional pinning centers may generate more frequent interruptions in domain wall motion. Interestingly, while the number of peaks in the BN signal remains constant with LF, the RMS captures relevant modifications in the intensity and temporal distribution of the Barkhausen events. In particular, an increase in average peak intensity, together with more energetic depinning events and localized clustering of peaks, enhances the overall signal power. The presence of irradiation-induced defects, such as CRP, plays a decisive role in this process by altering magnetization dynamics and producing larger Barkhausen jumps. These results highlight RMS as a highly sensitive parameter for detecting microstructural changes induced by irradiation, capable of capturing the evolving interaction between domain walls and defects beyond simple peak counting.

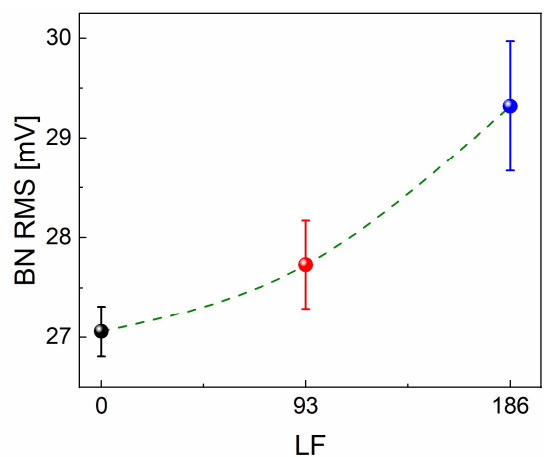

Figure 5: RMS value of Barkhausen noise (BN) as a function of the lead factor (LF).

## 4. Conclusions

In this work, we studied the changes induced by neutron irradiation in SA508 grade 3 steel using various magnetic techniques. The results show that the effects of irradiation are governed not only by neutron fluence, as expected, but also by the lead factor (LF). This indicates that, for the same fluence, the irradiation rate (or duration) strongly influences the distribution of copper-rich nanoprecipitates, significantly modifying the magnetic properties of the steel.

Our minor loops results, showed consistent variations $H_c^*$, $M_{rem}^*$, and $M_{max}$ across the three evaluated conditions. Both $H_c^*$ and $M_{rem}^*$ followed a similar trend, with an increase at low fields, a plateau at intermediate fields, and a decrease at higher fields. In contrast, $M_{max}$ exhibited a different trend: at low fields, irradiated samples showed higher magnetization, while at high fields, once saturation was reached, the non-irradiated sample displayed the largest magnetization. AC magnetic susceptibility measurements further revealed how irradiation and LF affect the dynamics of domain wall motion. The real component ($Re(\chi)$) increased with LF, indicating a higher density of smaller magnetic domains that can respond more easily to low applied fields. Meanwhile, the imaginary component ($Im(\chi)$) also increased with LF, reflecting the presence of more significant obstacles, such as nanoprecipitates, that hinder domain wall motion and enhance energy losses.

Barkhausen noise (BN) measurements confirmed the strong influence of LF on the magnetic response of SA508 steel. While the number of discrete Barkhausen jumps (peaks) remained nearly unchanged, the RMS value of the signal increased with LF. This reflects not only more energetic depinning events, but also changes in the temporal distribution of jumps, leading to higher overall signal power. The presence of irradiation-induced defects, such as copper-rich precipitates, plays a decisive role in this process by modifying magnetization dynamics and producing larger Barkhausen events. These features make RMS a particularly sensitive parameter for detecting microstructural changes induced by irradiation, and establish BN as a practical non-destructive technique, with strong potential for in-situ monitoring in nuclear environments.

In summary, the magnetic techniques employed in this study allow us to detect irradiation-induced changes in SA508 grade 3 steel safely and reproducibly. They are sensitive not only to fluence but also to LF, and their adaptability to nuclear environments makes them strong candidates for in-situ monitoring of the structural integrity of reactor pressure vessels.

## CRediT authorship contribution statement

**S. Passanante:** Investigation, Methodology, Formal Analysis, Writing - original draft, Writing – review and editing. **D. Goijman:** Investigation, Methodology, Formal analysis, Resources, Writing - review and editing. **M. R. Neyra Astudillo:** Investigation, Methodology, Formal analysis, Resources, Writing -review and editing. **C. D. Anello:** Investigation, Methodology, Formal analysis, Resources, Writing -review and editing. **R. Kempf:** Investigation, Methodology, Formal analysis, Resources, Writing -review and editing. **J. Milano:** Investigation, Methodology, Formal analysis, Resources, Writing -review and editing. **M. Gómez:** Investigation, Methodology, Formal analysis, Resources, Writing -review and editing. **J. Sacanell:** Investigation, Formal analysis, Supervision, Conceptualization, Methodology, Resources, Writing - original draft, Writing - review and editing, Project administration, Funding acquisition.

## Acknowledgements

The authors acknowledge financial support received from ANPCyT (PICT 2021-00495, PICT 2018-02397). The authors thank the RA-1 reactor operators and División de daño por radiación - CNEA for their assistance with the manipulation and cutting of the samples, especially to Gastón Romero and Axel Fariña Capra.

**Declaration of competing interest**

The authors declare that they have no known competing financial interests or personal relationships that could have appeared to influence the work reported in this paper.